\documentclass[pra,aps,final,twocolumn,nofootinbib]{revtex4-1}
\usepackage{epsfig}
\usepackage{latexsym}
\usepackage{xspace}
\usepackage{hyperref}
\usepackage[latin2]{inputenc}
\usepackage{indentfirst}
\usepackage{enumerate}
\usepackage{color}
\usepackage{colordvi}

\usepackage{amsmath}
\usepackage{amssymb}
\usepackage[english]{babel}
\usepackage{url}

\def\be{\begin{equation}}
\def\ee{\end{equation}}
\def\bea{\begin{eqnarray}}
\def\eea{\end{eqnarray}}
\def\l{\label}

\def\p{{\bf p}}

\def\r{{\bf r}}

\def\k{{\bf k}}

\def\d{\mbox{d}}

\def\siml{\;\hbox{\kern.1em \lower.7ex \hbox{$\sim$} \kern-1.12em
 \raise.5ex \hbox{$<$} \kern.1em}}
\def\simg{\;\hbox{\kern.1em \lower.7ex \hbox{$\sim$} \kern-1.12em
 \raise.5ex \hbox{$>$} \kern.1em}}

\begin{document}

\title{Viscosity of a classical gas: The rare-collision versus the
frequent-collision regime}

\author{A.G. Magner\thanks{magner@kinr.kiev.ua}}
\affiliation{\it  Institute for Nuclear Research, 03680 Kyiv, Ukraine}

 \author{M.I. Gorenstein}
\affiliation{Bogolyubov Institute for Theoretical Physics, 03680,
Kiev, Ukraine}

\author{U.V. Grygoriev}
\affiliation{Taras Shevchenko National University 03022 Kiev, Ukraine}

\begin{abstract}

The shear viscosity $\eta$ for a dilute classical gas
of hard-sphere particles is calculated
by solving the Boltzmann kinetic equation
in terms of the weakly absorbed
plane waves.
For the rare-collision regime, the viscosity $\eta$
as a function of the equilibrium gas parameters --
temperature
$T$, particle number density $n$,
particle mass $m$, and hard-core particle diameter  $d$ --
is quite different from that of the frequent-collision regime, e.g.\ ,
from the well-known result
of Chapman and Enskog.
An important property of the rare-collision regime is the dependence of
$\eta$ on the external ("non-equilibrium")
parameter $\omega$, frequency of the sound plane wave, that is
absent in the frequent-collision regime at leading order of the corresponding
perturbation  expansion.
A transition from the frequent to the rare-collision regime takes place when
the dimensionless parameter
$nd^2 (T/m)^{1/2} \omega^{-1}$ goes to zero.
\end{abstract}

\maketitle
\section{ Introduction }

As well known, the transport coefficients,
such as the thermal conductivity, diffusion, and viscosity,
can be defined as the linear responses on
small perturbations
of equilibrium systems \cite{kubo,zubarev,hofmann,wiranata-jp-14}.
Let us consider a classical gas system of hard-sphere particles.
An important
quantity, crucial for the
 description of all
transport properties of this system, is the particle mean-free path estimated
as
$l \sim 1/(n\,\pi d^2)\;$,
where $n$ is the particle number density and $d$ is
the hard-core particle diameter. This is an internal property
of the equilibrium system.
For external (dynamical) perturbations, one has to
introduce another scale $L$ which is the size of the
space region where the gas properties (e.g., temperature,
mean particle density, and collective velocity)
undergo essential changes.
Usually, the inequality, $l\ll L$, is satisfied.
This corresponds to the so-called frequent-collision regime (FCR), and
the transport coefficients can be calculated
as a perturbation expansion
over a small parameter $l/L$
(see, e.g., Refs.\ \cite{chapman,kogan,fertziger,silin,LPv10}).
The leading terms of these expansions are found
to be independent of the scale $L$.
For example, the shear viscosity $\eta$, calculated by Maxwell within concepts
of the molecular kinetic theory,
reads $\eta \sim n\;m\;v_T\;l$,
where $m$ is the particle mass and $v_T= \sqrt{2T/m}$ is the
thermal particle velocity\footnote{We use the units where the Boltzmann
constant is $\kappa^{}_B=1\;$.}. Thus, in the FCR,
$\eta$ is a function of only internal parameters of the equilibrium gas.
Since $l\propto 1/(nd^2)$,
one finds $\eta\sim \sqrt{mT}/d^2$
in the FCR, i.e., the shear viscosity
is independent of particle
number density at the leading order in
$l/L$.  Similar results are valid
for other transport coefficients in the FCR.

Accurate expressions for the
transport
coefficients in a gas of
classical hard-sphere particles
were obtained by Chapman and Enskog (CE) by using
the Boltzmann kinetic equation (BKE) and ideal hydrodynamic equations
for calculations of a time evolution in terms of  almost
the local-equilibrium distribution function within the FCR
\cite{chapman}.
The following expression for  $\eta$ was obtained
\cite{chapman,kogan,silin,fertziger,LPv10}:
\be\l{etaCE}
\eta_{\rm CE}=\frac{5}{16\sqrt{\pi}}\;\frac{\sqrt{mT}}{d^2}\;.
\ee
This
result was extended to a multicomponent hadron gas
in Ref.\ \cite{gorenstein}.
Several investigations
were devoted to go beyond the standard approach; see, e.g.,
Refs.\ \cite{abrkhal,GrJa59,Gr65,Si65,BrFe66,brooksyk,balescu,Le78,Wo79,Wo80,relkinbook,SGnpa85,Le89,prakash-pr93,baympeth,kolmagpl,magkohofsh,Ha01,pethsmith02,heipethrev,spiegel_BVKE-visc_03,smith05,PCprc11,AWprc12,AWcejp12,MGGPprc16,review}.
In the recent paper \cite{MGGPprc16}, the shear viscosity $\eta$
was calculated for a gas of particles
with both the short-range repulsive and long-range attractive
interactions described by the van der Waals equation of state \cite{marik2}.
This was  realized within the FCR in terms of a
strong suppression of the damping plane waves.

Many theoretical
\cite{Le78,Wo79,Wo80,relkinbook,Le89,Ha01}
and experimental \cite{Gr56,Me57,Gr65} investigations were devoted to
a weak absorption of the sound wave in  dilute gases.
For the sound absorption
coefficient $\gamma$, one can use the famous Stokes expression in terms
of the viscosity and thermal conductivity coefficients
(see, e.g., Refs.\ \cite{LLv6,Ha01}). Most results
for the absorption coefficient $\gamma$
were also obtained in the FCR.

Much less attention was paid to the rare-collision regime (RCR)
which takes place
at $l\gg L$; i.e., the so-called Knudsen parameter, $l/L$,
becomes large (see, e.g., Refs.\ \cite{kogan,silin,fertziger}).
Different analytical methods \cite{Wo79,Wo80,Ha01}
and numerical Monte Carlo simulations \cite{Ha01}
(see also the textbook \cite{kogan}) were used to calculate
the absorption coefficient $\gamma$
       by solving approximately the BKE.

The conditions of the RCR can be fulfilled in different ways.
For a small particle number density, a typical RCR situation
arises because of the finite system size. Note that for
$nd^3\rightarrow 0$
the mean-free path behaves as $l\rightarrow \infty$.
Therefore, Eq.\ (\ref{etaCE}) fails
for any finite physical system in the limits, $n\rightarrow 0$ and/or
$d\rightarrow 0$, where a density
and/or a diameter of particles is vanishing.
In both these limits, one has $nd^3\ll 1$ and,
thus, one results in the relationship
$l\gg L$. Thus, the FCR
is transformed to the RCR; and Eq.\ (\ref{etaCE}), obtained within the FCR,
becomes invalid.  Such a situation always appears for a gas expanding to
a free space:
the linear
size of the system $L$ increases due to a gas expansion, but the
mean-free
path $l\propto n^{-1}\propto L^3$
increases much faster. Thus, the RCR always takes place at the
latest stages
of the gas expansion to a free space.

The purpose of the present paper is to develop a general
perturbation method which can be applied in both
the FCR and the RCR  for the weakly absorbed plane-wave (WAPW)
solutions of the BKE.
The WAPWs play an important role in both
the theoretical studies and physical applications. They correspond to the plane
waves with a given frequency $\omega$ for which the amplitude only
slightly decreases
within the wave length $\lambda$.  The WAPW can take place within both the
FCR and RCR.
Thus, the perturbation expansion can be developed for small and large
parameter $l/\lambda$
for the FCR and RCR, respectively.

The FCR and RCR can be defined in alternative terms. Let us
consider the sound plane-wave propagation in the
infinite gas system. In this case ($L \sim \lambda$),
different regimes take place
because of different relationships between the mean-free path
$l$ and the wavelength $\lambda$ of the
propagating plane wave:
$l\ll \lambda$ and $l\gg \lambda$ correspond to the FCR and RCR,
respectively \cite{silin,Gr65,relkinbook}.
Introducing the wave frequency $\omega$ and frequency of two-particle
collisions $\tau^{-1}\sim v_T/l$, one finds another equivalent
classification for different collision regimes:
$\omega\tau\ll 1$
corresponds to the FCR,
and $\omega\tau\gg 1$ to the RCR.
A dimensionless quantity  $\omega\tau$ plays  the same role as
the Knudsen parameter mentioned above.
Small (large) values of $\omega\tau$  correspond to
the collision-term (inertial-terms) dominance in the BKE.
Important fields of the RCR applications are the ultrasonical absorption
(see, e.g.,  Refs.\ \cite{Gr65,bhatia,spiegel_Ultrasonic_01}
and special phenomena in the electronic plasma
\cite{silin}.
In the FCR we may use the perturbation expansion in
power series over $\omega\tau$ whereas in the RCR the parameter $\omega\tau$
becomes large, and the perturbation
expansion over $1/(\omega\tau)$  should be successful.
Therefore, in
these two collision regimes, one can expect a
different dependence of $\eta$
on the equilibrium gas parameters, $n$ and $T$, and proper particle
parameters, $d$ and $m$. In addition,
a dependence of $\eta$ on  external ("non-equilibrium")
parameter $\omega$, absent in Eq.\ (\ref{etaCE}) for the FCR  at the leading
order of the perturbation approach,
is expected
to appear
in the RCR. To our knowledge,
no  explicit analytical expressions for $\eta$
were so far presented for
the RCR. For completeness,
we
present also
the viscosity for a small absorption within the FCR
using the same perturbation method
but in
small parameter $\omega\tau \ll 1$.

The paper is organized as follows.
In Sec.\ II, the BKE approach with the relaxation time approximation for
the collision term through the WAPW is
discussed. In Secs.\ III and IV, derivations of the
expressions for $\eta$  in the FCR and RCR are presented, respectively.
The obtained results are used to calculate
the scaled absorption coefficient for
a sound wave
 propagation in the FCR and RCR. Section V is devoted to the discussions of
the results.
Section VI summarizes the paper, and Appendixes \ref{appA} and \ref{appB} show
some details of our calculations.

\section{
 The kinetic approach in a relaxation time approximation}\label{relaxation}

For a classical system of hard spheres, the single-particle distribution
function $f(\r,\p,t)$, where  $\r$, $\p$, and $t$ are the
particle phase-space coordinates, and the time variable, respectively,
is assumed to satisfy the BKE. The global equilibrium of
this system can be described by
the Maxwell
distribution as a function of the modulus of the particle momentum
$\p$ ($p\equiv |\p|$):
\be\l{maxwell}
f_{\rm eq}(p)=
\frac{n}{(2\pi mT)^{3/2}} \exp\left(-\;\frac{p^2}{2mT}\right)\;.
\ee
The particle number density $n$ and temperature $T$
are constants independent of the spacial coordinates $\r$ and time $t$.
For dynamical variations
of the equilibrium distribution (\ref{maxwell}),
$\delta f(\r,\p,t)=f-f_{\rm eq}$,
one obtains at $|\delta f|/f_{\rm eq}\ll 1$ the BKE
linearized over $\delta f$,
\be\l{Boltzlin}
\frac{\partial \delta f}{\partial t}+
\frac{{\bf p}}{m} \frac{\partial \delta f}{\partial {\bf r}}
= \delta St\;.
\ee
The standard form of the Boltzmann collision integral $\delta St~$
is used for hard spherical particles
\cite{chapman,silin,LPv10,MGGPprc16}.
For simplicity,  the attractive long-range
interactions discussed in
Ref.\  \cite{MGGPprc16}
are discarded.

In line with Refs.\ \cite{brooksyk,kolmagpl,review,MGGPprc16}, the
solutions of
Eq.\ (\ref{Boltzlin}) for $\delta f(\r,\p,t)$
can be sought in terms of the
WAPW,
\be\l{planewavesol}
\!\delta f(\r,\p,t)=
f_{\rm eq}(p) A(\hat{p})
\exp\left(-i \omega t + i k z\right),
\ee
where  $A(\hat{p})$ is a yet unknown function of the
angles, $\hat{p}\equiv \p \cdot \k/(pk)$. Here, $\omega$ and $\k$ are,
respectively, the frequency and a wave vector  of the
WAPW directed along the $z$ axis
($k=|\k|$). For convenience, the spherical phase-space
coordinates
with the polar axis directed to the
unit wave vector
$\k/k$
can be used.
The quantities $\omega$ and $k$
are connected by the equation
\be\l{freqom}
\omega=k\; c\; v_T =k\;c\;\sqrt{2T/m}\;.
\ee
In Eq.\ (\ref{freqom}), $\omega$ is
a given real frequency,
whereas a dimensionless sound velocity
$c$  and wave number $k$ are presented,
in general, as
complex numbers
\be\l{k}
c=c_r+i \;c_i\;,\qquad k = k_r +i\;\gamma\;.
\ee
A parameter $\gamma$ denotes
the absorption coefficient.
The imaginary quantities in Eq.\ (\ref{k})
are responsible for
a description of the dissipative process.

We use the standard definition of the
shear viscosity \cite{LLv6,MGGPprc16},
\be\l{etadef}
\eta =
\frac34 {\rm Re} \frac{\delta \sigma_{zz}}{\partial u_z/\partial z}\;,
\ee
through the
dynamical components of the stress tensor,
\be\l{dsigmadef}
\delta \sigma_{zz}=\int d \p\;
\left(\frac{p^2-3p_z^2}{3m}\right)\; \delta f(\r,\p,t)\;,
\ee
and the $z$ component of the collective velocity,
\be\l{udf}
u_z=\frac{1}{n}\;\int d \p\; \frac{p_z}{m}\;\delta f(\r,\p,t)\;.
\ee

In what follows, it will be convenient to expand the plane-wave amplitude
$A(\hat{\p})$
in Eq.\ (\ref{planewavesol}) over the spherical harmonics $Y_{\ell 0}(\hat{p})$
\cite{varshalovich}
\be\l{suml}
A(\hat{p})=\sum_{\ell=0}^{\infty} A_\ell
\;Y_{\ell 0}(\hat{p})\;.
\ee
To solve uniformly the BKE (\ref{Boltzlin}) in both the FCR and RCR, the
integral collision term $\delta St $
will be expressed in the form of the relaxation time
approximation
\cite{abrkhal,brooksyk,SGnpa85,heipethrev,kolmagpl,magkohofsh,spiegel_Ultrasonic_01,PCprc11,AWcejp12,AWcejp12,MGGPprc16,review}:
\be\l{tauapprox}
\delta St \approx
-\; \frac{1}{\tau}\;\sum_{\ell\ge 2}^{\infty} \delta f_\ell\;,
\ee
with the
relaxation time
\cite{MGGPprc16},
\be\l{tau}
\tau 
\sim \frac{1}{ n \;v_T \sigma}\;,
\ee
where $\sigma$ is the cross section for the two-particle collisions,
which is
given by $\sigma=\pi d^2$ for the case of the hard-sphere particles
of the diameter $d$.
Note that $\tau\sim l/v_T$
determines (up to
a constant
factor) the average time
of the motion between successive
collisions of the particles, and thus, $1/\tau$ is
approximately the collision frequency
in the molecular kinetic theory \cite{LPv10}. The summation over $\ell$
in Eq.\ (\ref{tauapprox}) starts from $\ell=2$ to obey the particle
number and momentum conservation in two-body collisions
\cite{brooksyk,kolmagpl,magkohofsh,review,MGGPprc16}.
The shear viscosity $\eta$ [Eq.\ (\ref{etadef})]
can be calculated analytically in the
two limiting cases:
$\omega\tau \ll 1$ (FCR) and
$\omega\tau \gg 1$ (RCR).

\section{Frequent collisions
}\label{sec-FCR}

In the FCR, the dispersion equation
for a dimensionless velocity $c$ reads
(see Appendix \ref{appA}
and Ref.\ \cite{MGGPprc16})
\be\l{dispeqFC}
c\left[c^2\left(1 + \frac{i}{\omega\tau}\right)
-\frac{3}{5} - \frac{i}{3 \omega\tau}\right] =0\;.
\ee
For nonzero solutions,
$c \neq 0$, at first order
in Knudsen parameter, $\omega\tau \ll 1$, one approximately finds
from Eq.\ (\ref{dispeqFC}),
\be\l{solFCR}
c_r = \frac{1}{\sqrt{3}}\;,\qquad c_i =
 - \frac{2\omega\tau}{5 \sqrt{3}}\;.
\ee
According to Eq.\ (\ref{k}) at first order of the perturbation
expansion, one then has
\be\l{krdef}
k_r =\frac{\omega}{c_r v^{}_{T}}\;,\qquad
\frac{\gamma}{k_r} = -\frac{c_i}{c_r}\;.
\ee
Therefore, for the scaled
absorption coefficient $\gamma/k_r$ in the FCR, one obtains
\be\l{sacFCR}
\frac{\gamma^{}_{\rm FC}}{k_r}= \frac25\;
\omega\tau\;.
\ee
Thus, for the scaled
absorption coefficient $\gamma/k_r$ in the FCR, one obtains
(Appendix \ref{appA})
\be\l{etaFCR}
\eta^{}_{\rm FC} = \frac{3\sqrt{\pi}\;nT}{10\; \omega}\; \omega\tau
=\frac{\sqrt{2\pi mT}}{10 \sigma}\;,
\ee
where $\sigma=\pi d^2$ is
the same cross section for
collisions of two hard-core spherical particles as
in Eq.\ (\ref{tau}). In Eq.\ (\ref{etaFCR}),
 the frequency $\omega $
is canceled at leading first-order
perturbation expansion over $\omega\tau$.
The shear viscosity $\eta^{}_{\rm FC}$ in the FCR behaves always as
$\eta^{}_{\rm FC}\propto \sqrt{mT}/\sigma$.
However, the numerical factor in this formula appears to be different for
different physical processes. For the strongly suppressed plane waves
(see Ref.\ \cite{MGGPprc16}), one has
 $|c_i| \gg |c_r|$, and the shear
viscosity approaches that in Eq.\
(\ref{etaCE}). For the WAPW in the FCR,
one finds another relation $|c_i|\ll |c_r|$
which leads to a different numerical factor in $\eta^{}_{\rm FC}$
[Eq.\ (\ref{etaFCR})] as compared to Eq.\
(\ref{etaCE}).

\section{Rare collisions}\label{sec-RCR}

Within the RCR, one can use the perturbation expansion over a small parameter
$1/(\omega \tau) \ll 1$.
After the substitution
of Eqs.\ (\ref{planewavesol})
and (\ref{tauapprox})
into Eq.\ (\ref{Boltzlin}) we derive
the RCR dispersion equation
for the WAPW sound velocity $c$
(see Appendix \ref{appB}),
\be\l{dispeqRC}
1 - \frac{i c}{\xi \omega\tau} - \left[
\frac{i\;c(3\xi^2+1)}{\xi\omega\tau}
-\epsilon_0\right] Q_1(\xi)  = 0\;,
\ee
where
$\xi=c[1+i/(\omega\tau)]$, $\epsilon^{}_0=+0$, and $ Q_1(\xi)$
is the Legendre function of a second kind given by Eq.\ (\ref{Q1}).
From Eqs.\ (\ref{dispeqRC}), (\ref{k}), and (\ref{krdef}),
one approximately
finds at $1/(\omega\tau)\ll 1$ quite a different solution as compared to
Eq.\ (\ref{solFCR}),
\be\l{RCRc}
c_r = 1,\qquad\qquad c_i=-\frac{1}{\omega \tau}\;.
\ee
Using Eqs.\ (\ref{freqom}), (\ref{k}), (\ref{krdef}), and (\ref{RCRc}),
one arrives at
\be\l{sacRCR}
k_r
= \frac{\omega}{v_T}\;,\qquad\qquad
\frac{\gamma^{}_{\rm RC}}{k_r}=\frac{1}{\omega\tau}\;.
\ee

For the shear viscosity (\ref{etadef}), one
straightforwardly obtains
(Appendix \ref{appB}):
\be\l{etaRCR}
\eta^{}_{\rm RC} = \frac{9\sqrt{\pi}\;n\;T}{4\omega^2 \;\tau}
= \frac{27\sqrt{2\pi}}{8}\;\frac{
n^2\; T^{3/2}\; \sigma}{\sqrt{m}\; \omega^2 }\;,
\ee
where $\sigma=\pi d^2$ as above.
 Notice that this viscosity is the leading first-order term of the
perturbation expansion over a small
parameter $1/(\omega\tau)$,
and therefore, $\eta \propto 1/\omega^2\tau$.
Comparing the expressions (\ref{etaRCR})
and (\ref{etaFCR})
for the shear viscosity,
one observes several distinct features of the RCR: 1) the dependence on
parameters $m$, $T$, and $\sigma$ in the RCR [Eq.\ (\ref{etaRCR})]
is completely different from that
in the FCR [Eq.\ (\ref{etaFCR})];
2) a dependence of $\eta$ on  particle
number density
$n$ appears in the RCR and
it is absent in the FCR; 3) a dependence of $\eta^{}_{\rm RC} $ on the
external
parameter $\omega$ exists
in the RCR [Eq.\ (\ref{etaRCR})]
whereas $\eta^{}_{\rm FC}$
 [Eq.\ (\ref{etaFCR})]
depends only on the internal gas properties
and is independent of $\omega$.

\section{Discussion of the results}
\l{sec-disc}

Let us compare the expressions for the shear viscosity in the RCR
[Eq.\ (\ref{etaRCR})]
and FCR [Eq.\ (\ref{etaFCR})].
From this comparison, one observes quite a
different dependence of $\eta^{}_{\rm RC}$ and
$\eta^{}_{\rm FC}$
on the quantities $n,T,d$, and $m$ which describe the
equilibrium classical gas.
Most striking is the
difference between two regimes in the limit of point like
particles $d\rightarrow 0$.
In this limit, $\eta^{}_{\rm FC}\rightarrow \infty$, whereas
$\eta^{}_{\rm RC}\rightarrow 0$.
As even a more remarkable property,
the difference between
the RCR and the FCR  is that the leading
[first-order in $1/(\omega\tau)$] term of $\eta^{}_{\rm RC}$ [Eq.\ (\ref{etaRCR})]
depends on a frequency $\omega$,
$\eta^{}_{\rm RC} \propto 1/\omega^2$, while
$\eta^{}_{\rm FC}$  (at the same first order but in
$\omega\tau$) is independent of $\omega$.

The definition of $\delta f$ given by Eq.(\ref{planewavesol})
is the same for both
  FCR and RCR. However, as explained in Appendixes \ref{appA} and
\ref{appB}, one has essentially different
  dispersion equations --
(\ref{dispeqFC}) in the FCR and
(\ref{dispeqRC})
in the RCR -- and their solutions $c$
for the wave velocity
[cf.\ Eqs.\ (\ref{solFCR})
and (\ref{RCRc})].
   As the result, from the same Eqs.\  (\ref{etadef})--(\ref{udf}),
one obtains  different
   final expressions: Eqs.\ (\ref{etaFCR})
for $\eta^{}_{\rm FC}$ and (\ref{etaRCR})
for $\eta^{}_{\rm RC}$.
 As we consider only the leading terms
 in the corresponding perturbation expansions for both the RCR and FCR,
 these expressions are valid for dilute gases.
 This is a general feature of the kinetic approach.
 Adding higher  order terms, one can hope to reach a wider range of
     applicability, including particularly  gases at higher density.

According to Eq.\ (\ref{sacRCR}), the dimensionless (scaled)
absorption coefficient $\gamma/k_r$ has a simple universal behavior in the
RCR
[$1/(\omega\tau)\ll 1$].
This dependence is very different from that of Eq.\ (\ref{sacFCR}) in the
FCR ($\omega\tau \ll 1$).

\begin{figure}
\begin{center}
\includegraphics[width=0.48\textwidth]{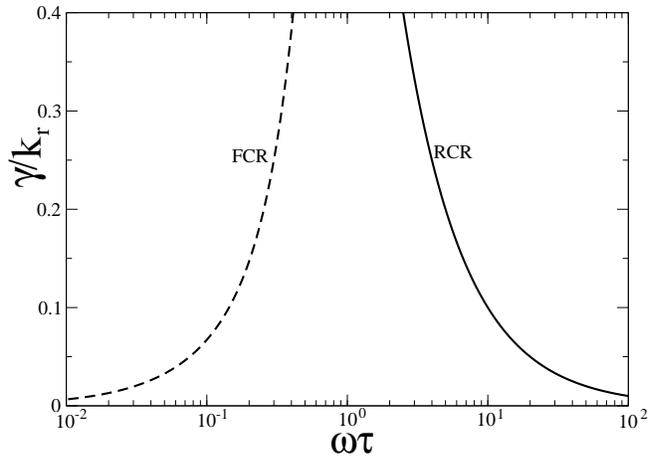}
\end{center}
\caption{
The scaled absorption coefficient $\gamma/k_r$ [Eqs.\ (\ref{sacRCR}) and
(\ref{sacFCR})] 
as a function of the Knudsen parameter $\omega \tau$
for the FCR  ($\omega \tau\ll 1$, dashed line) and RCR
($\omega \tau\gg 1$, solid red line).
}
\label{fig1}
\end{figure}

Figure \ref{fig1} shows the scaled absorption coefficients
$\gamma/k_r$  in the FCR (dashed line) and RCR (ssolid line) as
a function of the Knudsen parameter $\omega\tau$.
At $\omega\tau\sim 1$ a dramatic change of the
absorption coefficient $\gamma/k_r$  as a function of the Knudsen parameter is 
expected:
$\gamma/k_r$ increases in the FCR like
$\omega\tau$
at small $\omega \tau \ll 1$, according to Eq.\ (\ref{sacFCR}),
and decreases as $1/(\omega\tau)$ at large
$\omega \tau \gg 1$ in the RCR.

Equations
(\ref{etaRCR})
for $\eta^{}_{\rm RC}$ and (\ref{etaFCR})
for $\eta^{}_{\rm FC}$
can be also
used within the Stokes formula
   (see, e.g., Ref.\
\cite{LLv6}) for a weak absorption of sound waves.
   Substituting these equations
   into the Stokes equation and neglecting the thermal conductivity,
 one finds an agreement
   with the results for $\gamma/k_r$, presented in Fig.\ \ref{fig1}.
Thus, the whole  picture looks self-consistent.

\section{Summary}\label{Sum}
The shear viscosity $\eta$
is derived for the
damping sound in terms of the plane waves,
spreading in a dilute equilibrium gas of
classical particles described by hard spheres.
In the rare-collision regime the leading order of the
perturbation expansion over parameter $1/(\omega \tau)\ll 1$
for the shear viscosity $\eta^{}_{\rm RC}$   is quite
different from the first-order result in the frequent-collision regime.
First, very different dependencies of $\eta^{}_{\rm FC}$ and $\eta^{}_{\rm RC}$
on the
internal (equilibrium) gas quantities $n,\;T,\;d$ and $m$  are found.
Second, a basic difference is that $\eta^{}_{\rm FC}$ is
independent of
the non-equilibrium (external) frequency $\omega$,  whereas
$\eta^{}_{\rm RC} \propto 1/\omega^2$.

For small and large values of the Knudsen parameter $\omega\tau$,
one finds the scaled
absorption coefficient $\gamma/k_r$
growing proportionally to
$\omega\tau$
at $\omega\tau \ll 1$ and
decreasing as $1/(\omega\tau)$ at  $\omega\tau \gg 1$,
as well in the Stokes approach mentioned above.
Therefore, one can predict a maximum of $\gamma/k_r$ for the transition
between these two
collision regimes at $\omega\tau \approx 1$.
Thermal conductivity calculations for both the FCR and RCR can be
done within a more general approach, e.g., the
 linear response theory for solving the BKE. Within this formalism,  one can
 formulate the extended presentation for all kinetic coefficients
 suitable for their calculations in the nonperturbative region, too.
We plan to consider these problems in the forthcoming publications.
The results for the kinetic coefficients can be improved
by accounting for higher order terms in the perturbation
expansions and numerical calculations, towards the range of
 $\omega \tau$ close to one.
Our analytical results in the rare collision regime are universal and
do not include any fitting parameters.  Their accuracy
increases with increasing of $\omega\tau$.
They can be extended to  more general interactions
between particles as well as,
with the help of the linear response theory, to other transport
coefficients such as the thermal
and electric conductivity,  and the diffusion coefficients.

\begin{acknowledgments}
We thank D.V.\ Anchishkin, O.A.\ Borisenko, K.A.\ Bugaev, S.B.\ Chernyshuk,
Yu.B.\ Gaididey, V.P.\ Gusynin,
V.M.\ Kolomietz, B.I.\ Lev, V.A.\ Plujko,
S.N.\ Reznik, A.I.\ Sanzhur, Yu.M.\ Sinyukov, A.G.\ Zagorodny, and
G.M.\ Zinovjev for
many fruitful discussions.
One of us (A.G.M.) is
very grateful for the financial support of the Program of Fundamental
Research  to develop further cooperation with
CERN and JINR ``Nuclear matter in extreme conditions''
by the Department of Nuclear Physics and Energy of National
Academy of Sciences of Ukraine, Grant No. CO-2-14/2017,
for kind hospitality during his working visit
to the Nagoya Institute of Technology, and
also the Japanese Society of Promotion of Sciences
for financial support, Grant No. S-14130.
The work of M.I.G. was supported
by the Program of Fundamental Research of the Department of
Physics and Astronomy of National Academy of Sciences of Ukraine.

\end{acknowledgments}

\appendix
\section{The frequent-collision regime
}\l{appA}

For the shear viscosity $\eta$ [Eq.\ (\ref{etadef})], one should
calculate
the mean-velocity $u_z$ [Eq.\ (\ref{udf})] and
the stress-tensor $\delta \sigma_{zz}$ [Eq.\ (\ref{dsigmadef})]
component. Using Eqs.\ (\ref{planewavesol}), (\ref{suml}) and (\ref{maxwell}),
one finds (see Ref.\ \cite{MGGPprc16})
\bea\l{eqdeffinB1}
\eta= \mbox{Re}\left(\frac{9i\sqrt{\pi}}{4 \sqrt{15}}\;\frac{nT c}{\omega}\;
\frac{A_2}{A_1}\right)\;.
\eea
For calculations of the ratio $A_2/A_1$ and derivations of the
dispersion equation
to obtain the velocity $c$ in the FCR, we
substitute the plane-wave solution (\ref{planewavesol}) for the distribution
function $\delta f$
with the multipole expansion
(\ref{suml}) for $A(\hat{p})$
into the BKE (\ref{Boltzlin}).
 After simple algebraic
transformations, one finally arrives
 at the following linear equations for $A_\ell$ \cite{MGGPprc16}:
\be\l{Boltzeqfin}
\sum_{\ell=0}^{\infty} \mathcal{B}_{\mathcal{L}\ell}(c)\;A_{\ell}=0\;,
\ee
where
\bea\l{Boltzeqcoef}
&\mathcal{B}^{}_{\mathcal{L}\ell}(c) \equiv
c \delta^{}_{\mathcal{L}\ell} - C_{\ell 1;\mathcal{L}}
\nonumber\\
&+
i \Upsilon\;\delta_{\ell \mathcal{L}}
(1-\delta_{\ell 0})(1-\delta_{\ell 1})\;.
\eea
Here, $\delta_{\ell \mathcal{L}}$ is the Kronecker symbol,
\be\l{Upsilon}
\Upsilon=\frac{c}{\omega\tau}\;,
\ee
\bea\l{angintclebsh}
C_{\ell 1;\mathcal{L}}&=&\sqrt{\frac{4 \pi}{3}} \;
\int \d \Omega_p\; Y_{\mathcal{L}0}(\hat{p})\;
Y_{10}(\hat{p})\;Y_{\ell 0}(\hat{p})\nonumber\\
&=&\sqrt{\frac{2 \ell+1}{2\mathcal{L}+1}}\;
\left(C_{\ell 0,10}^{\mathcal{L}0}\right)^2\;,
\eea
$C_{\ell 0,10}^{\mathcal{L}0}$ are the Clebsh-Gordan
coefficients \cite{varshalovich}.

To derive
the dispersion equation (\ref{dispeqFC}) for the ratio
$c=\omega/(k v^{}_{T})$
in the FCR and, then, calculate the amplitude ratio in
Eq.\ (\ref{eqdeffinB1}) for
the viscosity $\eta$,
one has to specify a small perturbation
parameter $\omega \tau$ in the perturbation expansion
of
$\delta f(\r,\p,t)$.
Then, in the FCR
(small $\omega\tau $),  one can truncate
the expansion
of $A(\hat{p})$ (\ref{suml}) over
spherical functions $Y_{\ell 0}(\hat{p})$,  and relatively, the
linear system of equations
(\ref{Boltzeqfin})
at the quadrupole
value of $\ell$,
$\ell \leq 2~$,
because of a fast convergence of the sum
(\ref{suml}) over $\ell~$ at $\omega\tau \ll 1$ \cite{brooksyk}.
Then, for the amplitude ratios $A_{\ell+1}/A_\ell$, one finds
from Eq.\ (\ref{Boltzeqfin}) (see Ref.\ \cite{MGGPprc16})
\bea
\frac{A_0}{A_1}&=& \frac{1}{\sqrt{3}\; c}\;,\l{amp01}\\
\frac{A_2}{A_1}&=& \frac{2}{\sqrt{15}\;
\left(c + i \Upsilon\right)}\l{amp21}\;,
\eea
where $\Upsilon$ is given by Eq.\ (\ref{Upsilon}).
Within the  FCR,
because of large
$\Upsilon$ [Eq.\ (\ref{Upsilon})], one notes the convergence of the
coefficients $A_\ell$ of the expansion in multipolarities
(\ref{suml})
[see Eq.\ (\ref{amp21}) and Refs.\ \cite{brooksyk,kolmagpl,MGGPprc16}].
Therefore,
from the zero determinant of the 3X3 matrix
at the quadrupole
value $\ell \leq 2$ and $\mathcal{L} \leq 2$
for nontrivial solutions of the truncated
system of Eq.\ (\ref{Boltzeqfin}), we derive the
cubic dispersion equation (\ref{dispeqFC})
(see Ref.\ \cite{MGGPprc16}).
Substituting the underdamped (WAPW)
solution (\ref{solFCR}) for the sound velocity
$c$,
from Eq.\ (\ref{eqdeffinB1})
one obtains
Eq.\ (\ref{etaFCR}) for the shear viscosity $\eta^{}_{\rm FC}$.

The volume viscosity $\zeta$ can be calculated in a similar way:
\be\l{volviscdef}
\zeta =\mbox{Re} \frac{\delta \mathcal{P}}{\partial u_z/\partial z} \;,
\ee
where
$u_z$ and $\delta \mathcal{P}$ are, respectively,
the mean velocity $u_z$ [Eq.\ (\ref{udf})] and the dynamical
variation of the isotropic kinetic pressure,
\be\l{presszeta}
\delta \mathcal{P}=\int \frac{p^2}{3m}\;d \p \;\delta f(\r,\p,t)\;.
\ee
Using the WAPW variations
$\delta f$
given by Eq.\ (\ref{planewavesol}) and multipolarity expansion
(\ref{suml}) ($\omega=k_r c_r v^{}_{T} $ is real), one finds
\be\l{zetaFC}
\zeta=\frac{\sqrt{3 \pi}\; nT}{2\omega}\;
\mbox{Re} \left(\frac{c A_0}{i A_1}\right)\;,
\ee
According to Eq.\ (\ref{amp01})
for $A_0/A_1$,
 the complex
sound velocity $c$ is canceled, and therefore,
for a weak plane-wave absorption, one obtains $\zeta=0$ (see also Refs.\
\cite{LPv10,Ha01}).

\section{The rare-collision regime
}\label{appB}

For the integral collision term $\delta St$ of the BKE (\ref{Boltzlin})
in the $\tau$ approximation, one writes \cite{magkohofsh}
\bea\l{eq2}
\delta St&=&-\frac{\delta f}{\tau}+\frac{1}{\tau}
\left[A_0 Y_{00}(\hat{p}) + A_1 Y_{10}(\hat{p})\right] \nonumber\\
&\times& f_{\rm eq}(p)\;
\exp(-i\omega t+i kz)\;.
\eea
We introduce now new notations,
$\hat{p}=\cos \theta = x\;$ and $\xi=c+i\Upsilon\;$ [see Eq.\ (\ref{Upsilon})].
Using the expansion (\ref{suml})
of the amplitude $A(\hat{p})$ over spherical functions
$Y_{\ell 0}(\hat{p})$  with their orthogonal properties, and the
explicit expressions for $Y_{00}$ and $Y_{10}$, one obtains
\be\l{solvarphi}
\!\!A(x) = -\frac{i}{\sqrt{4\pi} (x-\xi)}\left[\left(\Upsilon
-i x \epsilon^{}_0\right) A_0
+\sqrt{3}\; x \Upsilon A_1 \right].
\ee
 For convenience of calculations we introduced also
 $\epsilon^{}_0=+0$ as an infinitesimally small parameter.
Integrating over the spherical angles of
$d \Omega_{p}=\sin\theta d \theta d \varphi $
with the spherical functions
$Y_{\ell 0}(\hat{p})$,
one has
\bea\l{defparphil}
A_\ell& &= \int A(\hat{p}) Y_{\ell 0}\left(\hat{p}\right)
d \Omega_{p}\nonumber\\
&=&\sqrt{\pi (2\ell+1)} \int^1_{-1} A(x) P_\ell(x) d x\;,
\eea
where $ P_\ell(x) =(4\pi/(2 \ell+1)^{1/2} Y_{\ell 0}(\hat{p})$
is the standard Legendre polynomials \cite{varshalovich}.
Substituting Eq.\ (\ref{solvarphi}) into Eq.\ (\ref{defparphil}), one
finds
\bea\l{Aell}
A_\ell&=&i\sqrt{2\ell+1}\left\{\Upsilon A_0 Q_\ell(\xi)\right.\nonumber\\
&-&\left.\left(\sqrt{3}\Upsilon A_1-i \epsilon_0 A_0\right)
\left[\delta^{}_{\ell 0} -\xi Q_\ell(\xi)\right]\right\},
\eea
where
$Q_\ell(\xi)$
are the Legendre functions of second kind,
\be\l{Qell}
Q_\ell(\xi)=\frac12 \int \frac{P_\ell(x)d x}{\xi-x}\;,
\ee
in particular,
\be\l{Q1}
Q_1(\xi)=\frac{\xi}{2} \ln\left(\frac{\xi+1}{\xi-1}\right) -1\;.
\ee
These functions
obey
the recurrence equations,
\bea\l{Qellrec}
Q_1&=&\xi Q_0(\xi)-1,\nonumber\\
2 Q_2(\xi)&=&3\xi Q_1(\xi)-Q_0(\xi)\;, ...\;,
\eea
For $\ell=0$ and $1$ one gets from Eq.\ (\ref{Aell})
the following system of linear equations with respect to
$A_0$ and $A_1$:
\bea\l{varphi01}
&&\left[1-i \Upsilon Q_0(\xi)-\epsilon^{}_0 Q_1(\xi)\right] A_0 -
i \sqrt{3} \Upsilon Q_1(\xi) A_1 \!=\!0\;,\nonumber\\
&&i \!\sqrt{3}\left(\Upsilon \!-\! i \epsilon^{}_0 \xi\right) Q_1(\xi) A_0 -
\left[1 \!-\! 3i \xi \Upsilon Q_1(\xi)\right] A_1\!=\!0\;.
\eea
Nonzero solutions of this system of linear equations exist under the
condition of zero for its determinant. This leads to the dispersion equation
(\ref{dispeqRC}) for the sound velocity $c$ through $\xi$.

Using Eqs.\ (\ref{Aell}) and (\ref{Qellrec}), one has
\be\l{AAQQ}
\frac{A_\ell}{A_{\ell-1}}=\sqrt{\frac{2\ell+1}{2\ell-1}}\;
\frac{Q_\ell}{Q_{\ell-1}}\; \quad \mbox{for}\quad
\ell \geq 1\;.
\ee
With the help of
Eq.\ (\ref{eqdeffinB1}), we arrive at \cite{MGGPprc16}
\be\l{eqdeffinB2}
\eta =
{\rm Re}\left[\frac{3i \sqrt{\pi}}{4}\; \frac{nT c}{\omega}\;
\frac{Q_2(\xi)}{Q_1(\xi)}\right]\;.
\ee
Taking into account the recurrence equations [Eq. \ (\ref{Qellrec})],
one can re-write $Q_2/Q_1$
in terms of  $Q_1^{-1}(\xi)$ and $\xi$, which are given by the
RCR dispersion equation (\ref{dispeqRC}).
In the limit $\epsilon^{}_{0} \rightarrow +0$ at first order in
$\Upsilon \sim 1/(\omega\tau)$, one then finds from Eq.\ (\ref{dispeqRC})
\be\l{Q1sol}
\frac{1}{Q_1(\xi)} \approx i \Upsilon \left(3 \xi +\frac{1}{\xi}\right)
\approx \frac{4i}{\omega\tau}\;.
\ee
We used solution (\ref{RCRc}) of the sound velocity $c$ to
the dispersion equation (\ref{dispeqRC})
in the second expression, valid at first-order perturbation expansion over
$1/(\omega\tau)$. Using this expression and solution (\ref{RCRc})
for $c$, from Eq.\
 (\ref{eqdeffinB2}) for the RCR shear viscosity $\eta$, one obtains
Eq.\ (\ref{etaRCR}).

\end{document}